\documentclass[prd,aps,amsmath,superscriptaddress,twocolumn,amssymb,nofootinbib]{revtex4}

\usepackage{latexsym}
\usepackage{epsfig}
\usepackage{amssymb}
\newcommand{\lp}{\left(}
\newcommand{\rp}{\right)}

\newcommand{\for}{\mx{ for }}
\newcommand{\mx}{\mbox}
\newcommand{\LF}{\left(}
\newcommand{\RF}{\right)}
\newcommand{\LT}{\left[}
\newcommand{\RT}{\right]}

\newcommand{\Rd}{\right.}
\newcommand{\ra}{\rightarrow}
\newcommand{\Ra}{\Rightarrow}

\newcommand{\3}{\frac{1}{3}}

\newcommand{\6}{\frac{1}{6}}

\newcommand{\where}{\mx{ where }}

\newcommand{\cP}{{\cal P}}

\newcommand{\cO}{{\cal O}}

\newcommand{\del}{\partial}
\newcommand{\ba}{\begin{eqnarray}}
\newcommand{\ea}{\end{eqnarray}}
\newcommand{\be}{\begin{equation}}
\newcommand{\ee}{\end{equation}}

\newcommand{\ka}{\kappa}

\newcommand{\da}{\delta}
\newcommand{\la}{\lambda}

\newcommand{\sa}{\sigma}

\newcommand{\Ga}{\Gamma}

\newcommand{\La}{\Lambda}

\begin{document}

\title{Could our Universe have begun with $-\Lambda$?}


\begin{abstract}
In this paper we present an informal description of the Cyclic Inflation scenario which allows our universe to ``start'' with a negative potential energy, inflate, and then gracefully exit to a positive potential energy universe. We discuss how this scenario fares in comparison with the standard inflationary paradigm with respect to the classic cosmological puzzles associated with the horizon, flatness and isotropy of our current universe. We also discuss some of the most debilitating problems of cyclic cosmologies, Tolman's entropy problem, and the problem with the overproduction of blackholes. We also sketch the calculation of the primordial spectrum in these models and possible observable signatures. We end with a special focus on the exit mechanism where the universe can transition from the negative to a positive  potential region. The treatise is based on an ongoing collaboration between the authors and closely follows conference presentations given on the subject by TB.

\end{abstract}

\author{Tirthabir Biswas$^{1,2}$,~Tomi Koivisto$^{3}$, and Anupam Mazumdar $^{4}$ }
\affiliation{$^{1}$Dept. of Physics, Loyola University, New Orleans, LA 56302, and\\
$^{2}$Dept. of Physics and Astronomy, University of Minnesota, Minneapolis, MN 55455\\
$^{3}$Institute for Theoretical Physics and Spinoza Institute,  Utrecht, The Netherlands.\\
$^{4}$Physics Department, Lancaster University, Lancaster, LA1 4YB, UK}

\maketitle
\section{Introduction}

It is apparent that observational consistency requires our universe  to have a positive vacuum energy ``today''~\cite{Perlmutter:1997zf}. Inflation which can explain the large scale structure of the universe~\cite{Komatsu:2010fb} also
requires a source of positive vacuum energy density (for a review see~\cite{Mazumdar:2010sa}), but there is yet no known theoretical reasons why that should be so. In fact, string theory, the leading candidate for a consistent theory of quantum gravity, naturally predicts the existence of negative energy vacua, and it has been quite a challenge to find ways that may lead to positive vacuum energies in the string theory framework~\cite{Kachru:2003sx}.
In general, if one looks at the potential energy coming from all the moduli in any fundamental theory,  one would expect to have both negative and positive potential regions, possibly with several local minima's dispersed liberally. In String theory this picture is often referred to as the ``landscape''~\cite{Douglas:2006es}.

So, let us consider a ``small'' relatively smooth patch in the ``Early Universe''. Without worrying  about any preconceptions,  what quantities would we normally associate with such a patch? (a) Vacuum energy, $\La$, which could be positive or negative, (b) Spatial curvature, $\rho_k\sim a^{-2}$, which again could contribute positively (open) or negatively (closed) to the energy budget, (c) some massless degrees of freedom or radiation, $\rho_r\sim a^{-4}$, (d) perhaps some massive modes as well, $\rho_m\sim a^{-3}$, and (e) possibly small amounts of anisotropy, $\rho_a\sim a^{-6}$. One could also add energy coming from coherent evolutions of scalar fields, and depending upon the form of the potential energy the cosmological dynamics could be complicated. For simplicity however, we consider only two extreme cases:  massless scalars whose kinetic energy density behaves in a manner very similar to that of the anisotropies, $\rho_{\phi}\sim a^{-6}$, and very slowly varying potentials. In the former case the scalar energy can be clubbed with $\rho_a$, while the latter  can be accounted for by simply assuming $\La=\La(\phi)$. 

So what kind of cosmologies can one expect? Of course, it depends on the different parameters. If for instance, anisotropies are large, then one has to abandon FLRW cosmology all-together and one would typically be stuck in a chaotic Mixmaster behavior, not particularly conducive to create an universe such as ours. Hence, from the very beginning we assumed a relatively smooth patch where FLRW is a good approximation. Let us further look at the  expanding universe case, rather than the contracting one. If $\La>0$, then we end up with standard slow-roll inflation, all the different energy components listed above dilutes away, and within a few efolds $\La$ comes to dominate the show. This  potentially can lead to a viable inflationary scenario where the universe keeps expanding monotonically, passes through the usual phases of radiation and matter domination, and eventually comes to be governed by dark energy around today. This is of course the Standard Model of Cosmology, but what if $\La<0$~\footnote{The case when $\La\sim 0$ is rather interesting because this can lead to what is popularly known as an emergent universe~\cite{emergent1,emergent2}, but we are not going to discuss this possibility here.}. 

If $|\La|> \rho_r+\rho_m+\rho_k$, FRW evolution is inconsistent, most likely the universe would be stuck in a static anti-de Sitter like universe containing massless and massive excitations. Again, not the kind of universe we find ourselves in. So, let us look at the opposite case when $|\La|< \rho_r+\rho_m+\rho_k$. As the universe expands, again all the matter components dilute and eventually cancel the negative cosmological constant causing the universe to turnaround and start contracting. In GR, this leads to an eventual collapse, clearly not what   our universe experienced. Very few cosmologists believe in actual existence of a Big Crunch/Big Bang singularity. It is expected that the quantum theory of gravity would resolve such singularities. We don't know how it would be resolved, and also what kind of space-time, if at all, would emerge from this ``almost collapse''. There have been suggestions both from top-down~\cite{Bojowald:2006da,ashtekar,Shtanov}, and bottom-up~\cite{Freese,Baum} approaches that the universe would transition non-singularly from the phase of contraction to a phase of expansion via the so called ``Big Bounce''. Suppose that's what happened, suppose the universe bounces whenever some critical Planckian energy density is reached. What happens next? Well, the same story simply repeats itself. Matter will dilute and when it decreases enough that the negative cosmological constant can cancel it, the universe again turns around ensuring the cyclic pattern.

Can this evolution lead to an universe like ours? Naively, the answer is no, because $|\La|$ is typically expected to be large, given possibly by the string/GUT scale, a few orders of magnitude below the Planck scale. Then each of the cycles would only last a very short time $\tau\sim M_p/\sqrt{\La}\sim 10^{-33}$ s, much too short to be interesting. Essentially we will have a nonsingular periodic but uninhabitable cosmology. However, we did forget something; interactions between different species. Interactions generally create entropy which can only increase monotonically breaking the periodicity of the evolution. In fact, this was precisely what Tolman pointed out in the 1930's giving rise to Tolman's entropy problem~\cite{tolman1,tolman2}. However, this turns out to be a great savior for the scenario. As will be described in more details in the next section, also see~\cite{Biswas:2009fv,Biswas:2010si}, entropy tends to increase by the same factor in every cycle, while the time period of the cycles remain a constant since it is governed by $\La$. This means that the universe must be growing by the same factor in every cycle giving rise to an overall inflationary growth! Most of the advantages of standard slow-roll inflation is straight forwardly transferrable to this scenario, including the production of near scale-invariant density fluctuations~\cite{Biswas:2010si}. However, the scenario suffers from an obvious drawback.

After all we have been living in a positive potential  energy region for the last 13 billion years, and the universe have been expanding monotonically during all this time. Is there a way to go from
negative energy regions to positive ones, have we neglected something again? In  an expanding background it is known that energy densities can only decrease, and hence once the universe is in a negative energy phase, there is no way for it to claw back up to the positive region. However, in contracting phases the reverse is true {\it i.e.},  the  energy density  increases, and in this article we will argue that  contracting phases can indeed facilitate a transition from negative to positive potential energy regions once one realizes that the cosmological constant should really be thought of as the potential energy coming from all the scalars (moduli) in one's theory. Thus our universe could have been ``exploring'' the negative potential regions, when at some point  it made a jump to a positive potential energy region,  ushering in a monotonically expanding phase which we are currently inhabiting in.

The main focus of this presentation is the graceful exit mechanism. We emphasize that a critical assumption required to make this mechanism work is that the universe (non-singularly) bounces from the contracting to the expanding phase at $\rho_{\bullet}$.
See for instance~\cite{bkm} for such a scenario involving non-local modifications of gravity. For technical simplicity, here we employ negative $\rho^2$ type of corrections~\cite{ashtekar,Bojowald:2006da,Shtanov,Lidsey:2006md}  in the Hubble equation to obtain our bounce.
From now on the subscript ``$\bullet$'' will refer to quantities at the bounce point.

We start by providing the ``big picture'' in the cyclic inflation scenario in section~\ref{tcis}. Next  in section \ref{toy}, we provide a qualitative description  of the graceful exit mechanismn. In section~\ref{numeric}, we study the mechanism numerically. Generalization and robustness of this mechanism is then investigated in a string landscape context in section~\ref{landscape}, and we conclude in section~\ref{sec:conclusions}.
\section{The Cyclic Inflation Scenario}
\label{tcis}
One of the motivations for considering cyclic cosmologies has been to provide an alternative to the standard monotonically expanding inflationary paradigm. In a series of papers~\cite{Biswas:2008ti,Biswas:2009fv}, a very different way of obtaining near scale-invariant fluctuations have been proposed in the context of  cyclic models. The main point is that in the context of cyclic cosmologies, if one considers a multi-fluid system where some components are allowed to fall out of thermal equilibrium, then in general, entropy will be produced through interaction of this species with the thermal bath. This dramatically alters the cosmology and opens up a completely new way of realizing inflation!

In order to illustrate the basics, let us consider a universe with a negative cosmological constant (CC). The Hubble equation corresponding to flat  Friedman-Lemaitre-Robertson-Walker (FLRW) cosmology reads
\be
3H^2=[\rho_m(a)-\La]/M_p^2\ ,
\label{hubble}
\ee
where $\rho_m$ represents any form of positive energy density source, such as radiation (massless excitations), kinetic energy of scalars, or non-relativistic matter. Consistent cosmology requires $\rho_m\geq \La$. In an expanding phase $\rho_m$ dilutes, and eventually it is canceled by the potential energy, $-\La$, when $H\ra 0$ signalling a ``turnaround'' to a contracting phase. Thereafter, $\rho_m$ increases, and according to our assumption once it reaches $\rho_{\bullet}$, the universe bounces back to an expanding phase. Thus, in the presence of matter a universe with a negative CC does not remain stuck in an anti-deSitter vacuum, but rather  ``cycles'' with a characteristic time period of $\tau\sim \sqrt{ M_p^2/\La}$! Of course, during the bounce phase the General Relativistic Hubble equation (\ref{hubble}) is not valid, but for the success of our mechanism  none of the details  of the bounce matter, only the energy scale is important. Also, the above cyclic evolution  does not depend on the kind of matter the universe has. This is in contrast to previously considered transition mechanisms which rely upon specific features of the models supposed to describe quantum gravity phase, such as nonlocals operators~\cite{Prokopec:2011ce}, inverse volume corrections in loop quantum cosmology \cite{Lidsey:2004ef,Mulryne:2004va,Nunes:2005ra,Mulryne:2005ef} or braneworld modifications~\cite{Lidsey:2006md}.

Next, let us say that $\rho_m$ consists of  ``small'' amounts  of a  non-relativistic matter (NR) species  along with relativistic degrees of freedom (radiation).  In this case, the universe ``turns around'' from a phase of expansion to a phase of contraction whenever  the total matter density (radiation + NR) equals $\La$, while it bounces back from the phase of contraction to the next phase of expansion whenever the critical Planckian bounce density, $\rho_{\bullet}$,  is reached. Importantly, this means that the cyclic evolution is periodic in terms of energy density.
To obtain an asymmetric (in scale-factor) cyclic evolution, one only has to require that the relativistic and the
non-relativistic species interact via scattering and decay processes; so, they remain in thermal equilibrium above a certain critical temperature, $T_c$, but  below this temperature the massive NR degrees of freedom fall out of equilibrium, and consequently when they decay into radiation,  thermal entropy is generated. During the contraction phase, once the temperature becomes greater than $T_c$, scattering processes become effective in replenishing the NR particles from radiation, and the system can return to thermal equilibrium before the start of the next cycle.

The details involving such processes have been studied using Boltzman equations in conjunction with the Hubble equation in~\cite{Biswas:2008ti,Biswas:2009fv}, more numerical investigations are currently underway~\cite{duhe}. Due to the periodic behavior of the various energy density components,  one crucially finds that the entropy increases by the same factor in every cycle. For instance, if $S_n$ denotes the entropy at the ``beginning'' of the $n$th cycle, then under some simplistic assumptions one can derive that
\be
\ka\approx  {1.2\mu\Ga M_p T_c g^{1/4}\over \La^{3\over4}}+\cO(\mu^2)\ , \ \where (1+3\ka)\equiv {S_{n+1}\over S_n} \ ,
\label{entropy}
\ee
where $\mu$ corresponds to the ratio of the equilibrium energy densities of the dust and radiation at $T_c$, $\Ga$ is the decay time,  and $g$ is the number of ``effective'' massless degrees of freedom in the radiation fluid.
Further, since the entropy is proportional to the volume this means that if $a_n$ refers to the scale factor at the bounce point of the $n$th  cycle, then ${a_{n+1}}/{a_n}\approx (1+\ka)^{1/3}$, a constant. Thus, over many asymmetric cycles the evolution of the universe resembles that of ordinary inflation with an average Hubble expansion rate $    H_{\mx{av}}\approx{\ka/\tau}$ for $\ka\ll1$, which turns out to be the phenomenologically relevant case.

As an immediate consequence,   this ``cyclic inflation" (CI) phase can address the usual cosmological puzzles such as isotropy, horizon, flatness and homogeneity, please see~\cite{Biswas:2009fv} for a detailed discussion.
Consider, for instance, the problem with the  growth of inhomogeneities, which is related to the problem of overproduction of black holes in cyclic models~\cite{blackholes}.  The matter fluctuations, $\da \equiv \da \rho/\rho$ can only grow as long as their wavelengths are larger than the Jean's length, $\la_J$,  given by:
$$ \la_J\sim c_s{M_p/\sqrt{\rho}}\mx{ where } c_s^2\equiv {\del p/\del \rho}$$
is the sound velocity square. Now, in our scenario the cycles are short, and most of the energy content is in radiation so that the sound speed is very close to the speed of light, and accordingly  $\la_J \sim \tau$. In other words, the ``sub-Hubble''\footnote{In the context of cyclic universes, by sub-Hubble fluctuations one really means fluctuations which are smaller than the cosmological time scale.} fluctuations don't grow.   On the other hand, once the wavelengths become  larger than the cosmological time scale, $\tau$, they become super-Hubble fluctuations and evolves according to the Poisson equation: $\da _k= {k^2\Phi_k/ (a^2\rho)}$, where $\Phi_k$ is the Newtonian potential characterizing the metric perturbations. Now in the super-Hubble phase, $\Phi_k$ becomes a constant while $\rho$ oscillates between a minimum and a maximum energy density. Thus we have: $\da _k< {k^2\Phi_k/( a^2\rho_{\min})}\sim  {k^2\Phi_kM_p/( a^2\sqrt{\La})}$. In other words $\da _k$ falls as $a^{-2}$ in the super Hubble phase just as in ordinary inflation.

Next, let us look at the problem  of getting embroiled in a chaotic Mixmaster  behavior as one approaches the ``Big Bounce''; since  anisotropies are known to grow as $\sim a^{-6}$,  cyclic cosmologies are notoriously plagued with this problem. However, in the CI model once the cyclic-inflationary phase is ``activated'' in a small and sufficiently smooth patch of the universe the chaotic Mixmaster behavior is avoided in subsequent cycles because the scale factor at the consecutive bounce points keeps growing and the universe becomes more and more isotropic. Very similar reasoning also resolves the flatness problem.

Moreover, just as in standard inflation, one expects to generate near scale-invariant perturbations during the CI phase. An easy way to see this is to recall that as long as the wavelengths of fluctuations are smaller than the cosmological time scale the power spectrum typically goes as  $\cP_{\Phi}\sim {\rho/ M_p^4}$, and this amplitude approximately freezes when the modes ``cross'' the Hubble radius. In the cyclic inflation scenario for $\ka\ll 1$, only a few modes make their ``last exit" into the superHubble phase in a given cycle. These exits therefore always occur near the turnaround of each of the cycles. If $V(\phi)$ is rolling very slowly, the turnaround energy density remains approximately constant throughout the entire cyclic inflationary phase leading to an approximately constant amplitude of perturbations. A more detailed argument is provided in~\cite{Biswas:2010si}.


While the CI scenario provides a new cosmological paradigm, it has to still contend with the graceful exit problem. If one is stuck in a $-$ve Cosmological constant, then the above inflationary phase persists forever and one can never obtain an universe like ours. The analysis that will be presented in the next couple of sections however will clearly demonstrate that if we have a scalar field rolling down a negative potential,  then there can indeed be a last cycle if the scalar field gets an opportunity to exit from the negative potential region to the positive potential region.

Finally, we need to revisit the issue of geodesic completeness in the context of cyclic inflation scenario. This is a  subtle  issue, if one tracks, say, the minimum of the oscillating space time, then one finds that it has the traditional inflationary trajectory, and the problem of past geodesic incompleteness becomes apparent~\footnote{In the context of cyclic universe, this is also known as Tolman's entropy problem.}! Fortunately, there appears a natural resolution that has been discussed in details in~\cite{Biswas:2008ti,Biswas:2009fv,Biswas:2010si}. For a closed universe, as one goes back in cycles, there comes a point when the curvature energy density becomes more important than the vacuum energy density. (Curvature density blue shifts as $a^{-2}$ with  decreasing scale factor, while the vacuum energy density just remains a constant.) Once this happens, the universe no longer turns around due to the negative vacuum energy density, but much earlier when the radiation density equals the negative curvature. Such a dynamics lead to an universe which undergoes progressively small oscillations around a constant scale factor as $t\ra -\infty$~\cite{Biswas:2008ti}. The space-time is, in fact, very reminiscent of the emergent universe scenario advocated in~\cite{emergent1,emergent2}. Thus the complete cosmological evolution in these scenarios consists of four distinct phases: Phase I or the cyclic-emergent phase, Phase II or cyclic-inflationary phase, phase III or the graceful exit phase, and finally an everlasting expanding phase.
\section{The Exit, Analytical Arguments}
\label{toy}
We are now going to focus on the exit mechanism. If $\La$ was a constant we would have a cyclic evolution ``forever'' in the past and future, which  would be a rather uninteresting space-time although geodesically complete and free from singularities~\footnote{A space-time metric is said to be geodesically incomplete in the past if one has particle trajectories which end abruptly at a finite proper time in the past with the particles finding themselves no where to go (or more accurately no where to come from). If the scale factor vanishes as $t\ra -\infty$, that typically signals such a problem. For instance, this is what happens in ``past eternal'' inflationary space times where the scale factor go like $e^{\la t}$~\cite{Borde,Linde}. In the present model, as $t\ra -\infty$, the scale factor doesn't go to zero but rather oscillates between a finite minimum and a maximum corresponding to finite minima's ($\La$) and maxima's ($\rho_b$) in energy density. All the trajectories can be trivially continued to the infinite past.}.  Fortunately, in a realistic scenario one does not expect $\La$ to be a constant, but rather identifies it as the potential energy associated with the various scalar fields in the model. Since here we want to study whether the universe can move from negative to positive potential regions, let us consider a very simple scenario involving a single scalar field that has been rolling along essentially a very flat negative potential, and then experiences a steep incline which can ``potentially'' take it to the positive side. For the moment, let us ignore the tilt in the negative side of the potential so that approximately
\be \label{potential}
V(\phi)=\left\{
\begin{array}{cc}-\La & \for \phi>0\\
-\La-\mu^3\phi & \for \phi<0
\end{array}
\Rd
\ee
The potential reaches zero at $\phi_{0}\equiv -\La/\mu^3$. We want to find out  whether $\phi$ can get enough velocity in the ``last'' contracting phase so that it can climb up the positive side of the potential. What we will show is that even if the kinetic energy is very small at the turnaround of the ``last'' cycle, $K_{\circ}\ll \La$, it is possible for the universe to make the transition. From now on the symbol ``$\circ$'' will denote the turnaround point.

In general, we expect both massive and massless excitations to be present in the universe. Although our analysis goes through for most types of additional matter sources, for the purpose of illustration, we will take this to be radiation  with $\rho_{r\circ}\approx\La$. Now, as the universe contracts, the scalar field will start gaining kinetic energy. Since there is no potential gradient as long as $\phi>0$, we have
$K=K_{\circ}({a_{\circ}/ a})^{6}$.
As the kinetic energy increases, it becomes possible for the kinetic energy to start dominating over the negative potential energy. It is straight-forward to check that this happens as long as
\be
K_{\circ} \gtrsim {\La^{5\over 2}/ M_p^6}
\ee
which is a rather mild requirement since $\La$ is expected to be given by the string scale $\sim 10^{15}-10^{16}$ Gev, so that $ {\La^{3/ 2}/ M_p^6}\sim 10^{-18}$. The crucial point is that once the kinetic energy becomes larger than $\La$, the total scalar energy density becomes $+$ve and the scalar field can no longer turn back while in the negative potential region as long as the universe is contracting~\cite{Felder}.

Let us consider a special case when kinetic energy just cancels the negative potential energy at $\phi=0$. The general situation will be considered numerically. At this ``equality epoch'', labeled by ``$e$'', we have
\be
 a_e=a_{\circ}\LF{K_{\circ}/ \La}\RF^{1/6}\ \Ra \rho_{r,e}=\La\LF{\La/ K_{\circ}}\RF^{2/3}\ .
\ee

We are now going to study the second phase of evolution when the potential is rising. The Klein-Gordon equation in this phase reads
$\ddot{\phi}+3H\dot{\phi}=\mu^3$. Since, at the equality, the energy density is dominated by radiation, and the kinetic energy is equal to $\La$, we have
\be
3H\dot{\phi}|_e\sim \frac{\La}{ M_p}\left(\frac{\La}{ K_{\circ}}\right)^{1/3}
\ee
Thus if
\be
\mu^3\ll  \frac{\La}{M_p}\LF\frac{\La}{K_{\circ}}\RF^{1/3} \sim |3H\dot{\phi}|_e
\label{mumax}
\ee
the scalar field continues to behave as a free field. Note that $\mu$ remains a constant while the Hubble term keeps increasing during contraction, making the approximation progressively better. Now, our main goal is to find out whether $\phi$ can reach $\phi_{0}$ before the total energy density $\rho\approx \rho_r+K$ reaches the bounce density $\sim M_p^4$.

The Hubble equation
\be
H^2={1\over 3 M_p^2}\LT \rho_r+K\RT={1\over 3 M_p^2}\LT \rho_{r,e}\LF{a\over a_e}\RF^{-4}+\La\LF{a\over a_e}\RF^{-6}\RT\ ,
\ee
can be recast into
\be
a{d\phi\over da}=M_p\sqrt{6}\LF{a\over a_e}\RF^{-1}\LT\LF{\La\over K_{\circ}}\RF^{2/3}+\LF{a\over a_e}\RF^{-2}\RT^{-{1\over2}}
\label{hubble}
\ee
and solved exactly in terms of  $y\equiv (a/a_e)(\La/ K_{\circ})^{1/3}$:
\be
-\phi_{\bullet}=\sqrt{6}M_p\LT\tanh^{-1}\LF{1\over \sqrt{1+y_{\bullet}^2}}\RF-\tanh^{-1}\LF{1\over \sqrt{1+y_e^2}}\RF\RT
\ee
where $y_e=\LF{\La/ K_{\circ}}\RF^{1/3}$ and $y_{\bullet}=\LF{\La/ M_p^4}\RF^{\6}\LF{\La/ K_{\circ}}\RF^{\3}$
are the initial and final values of the variable $y$. In particular $y_{\bullet}$ was estimated by assuming that the bounce occurs when the kinetic energy density equals the Planckian bounce density. Indeed near the bounce one cannot trust the GR evolution (\ref{hubble}), but the new physics should only lead to $\cO(1)$ modifications as it's effect is expected to kick in only at Planckian densities. We have verified this numerically.

To check whether the scalar field can indeed cross the negative potential region let us compare $\phi_{0}$ and $\phi_{\bullet}$. Since $y_e\gg1$, and $y_{\bullet}\ll 1$, we have
\be
-\phi_{\bullet}\approx \sqrt{6}M_p\tanh^{-1}\LF{1-y_{\bullet}^2/2}\RF\sim \cO(10)M_p
\ee
Now, if we want to be consistent with our approximation (\ref{mumax}), the minimum value $\phi_0$ can take is
\be
-\phi_{0}=M_p\LF{K_{\circ}/ \La }\RF^{1/3}\ll M_p
\ee
Thus we see, it should be easily possible to gracefully exit the negative potential region!

Although in the present manuscript we have not introduced a second matter component and possible exchange of energy between it and the radiation fluid, the existence of such a fluid and their interaction will not play any significant role in the exit mechanism proposed. Firstly, in the CI paradigm, the NR matter component is always subdominant to radiation, and secondly, during the bounce phase as the universe contracts, the matter energy density becomes even more insignificant as compared to the kinetic energy of the scalar field and the radiation density.
\section{Numerical Analysis}
\label{numeric}

\begin{figure}
\includegraphics[width=9.75cm]{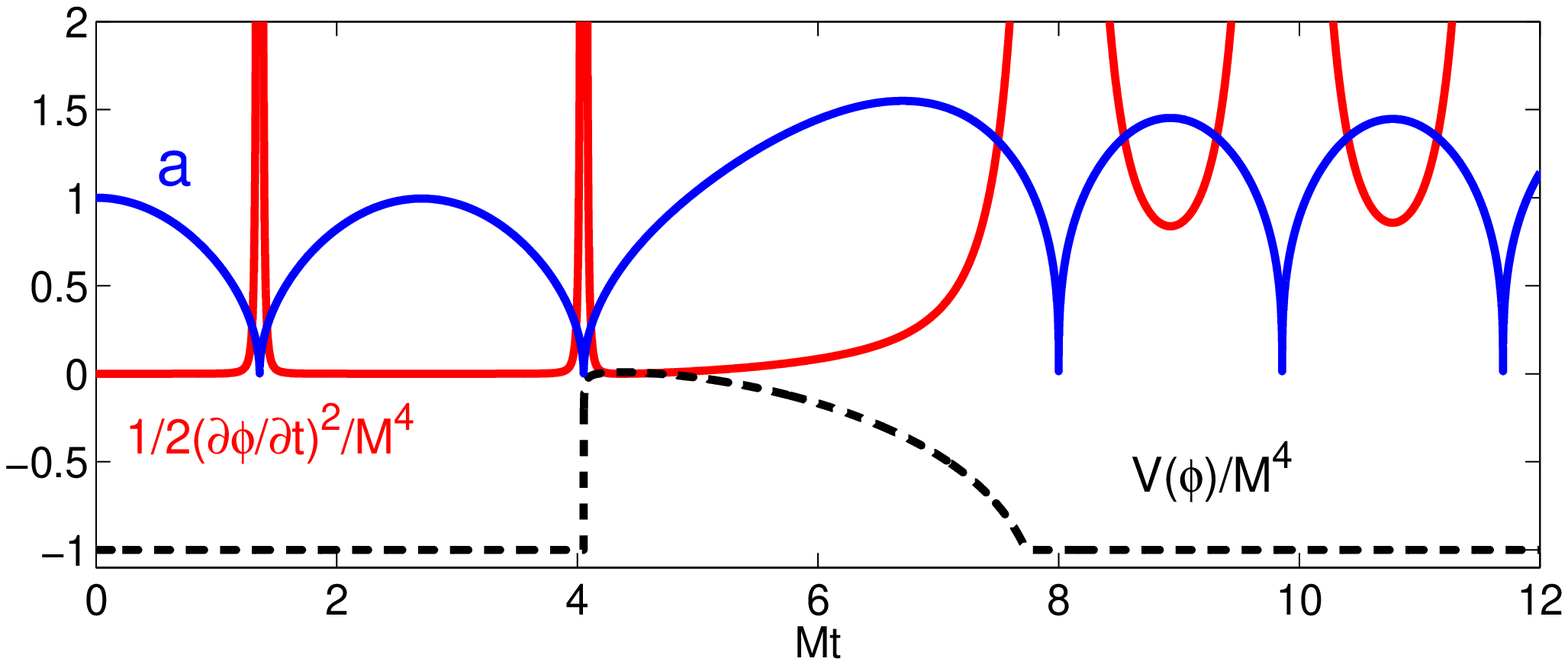}
\includegraphics[width=9.75cm]{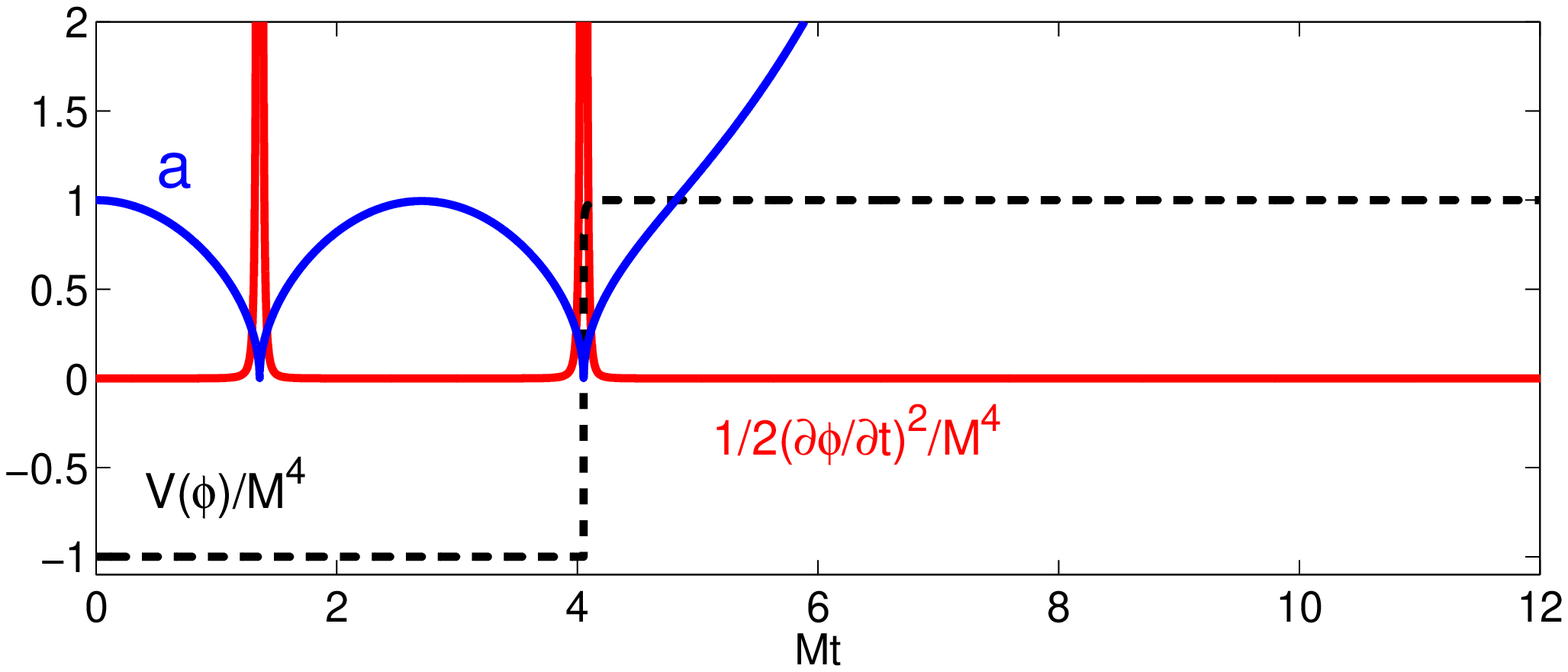}
\caption{\label{evolution}
Evolution of the kinetic and potential energies. The solid blue line shows the scale factor which is normalized to unity at the turnover at $t=0$.  Here we have set $\rho_{\bullet} = 10^9 M^4$, $\mu^3=M^3/2$ and $\La=M^4$.
{\bf Upper panel}: The field reaches the upward slope in the expanding phase. The force overcomes the friction which is diluting with the expansion, and consequently the field rolls back down the hill and returns to the cycling evolution.
{\bf Lower panel}: The field is released from a different position and crosses the upward slope completely in the bouncing phase. The potential becomes positive, and consequently the expanding phase continues unstopped. This is the more typical evolution: in fact one has to choose the initial conditions with some care for the field to end up in an evolution like the upper panel.}
\end{figure}

\begin{figure}
\includegraphics[width=9.75cm]{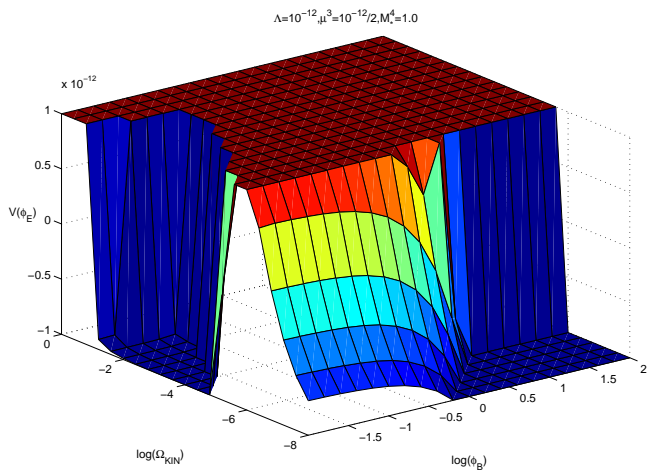}
\caption{\label{initial}
The effect of initial conditions for a choice of the three scales.
The parameters are in units of Planck mass.
x-axis: (10-based logarithm of) the initial value of the field at the turnover.
y-axis: (10-based logarithm of)  $K_{\circ}/\Lambda$. Thus, at y=0 there's no radiation but kinetic energy is equal to the magnitude of the cosmological constant. z-axis: $V(\phi)$ at the highest point it reaches. The red plateau is basically where we have graceful exits.}
\end{figure}

Although the mechanism proposed here does not depend on the details of the bounce, but only on the scale of the bounce, for numerical purposes we use the quadratic modification to the Hubble equation, $3H^2=8\pi G\rho[ 1-{\rho/\rho_{\bullet}}]$, as previously considered in~\cite{Freese,Baum}. The energy density $\rho$ consists, effectively, of radiation and a canonic scalar field with the potential (\ref{potential}).
A couple of typical evolutions are shown in Fig.~\ref{evolution}, while
Fig.~\ref{initial} exhibits the regions in the space of initial conditions where we obtain the graceful exit. Actually, for numerical studies we looked at situations where the field not only exits from the negative potential region, but is also able to jump to a height of, $V(\phi)\sim \La$.

The favorable conditions span large ranges of initial field values and the kinetic energy need not be  large initially. In Fig. \ref{initial} for instance, we see that even when the kinetic energy is  $10^{-8}$ times smaller than the potential energy, a successful exit is  possible with realistic parameter values. As the initial kinetic energy increases, it becomes more and more favorable for the universe to make the transition, and in particular when we have no radiation, the universe always makes a successful exit.

\begin{figure}
\includegraphics[width=8.75cm]{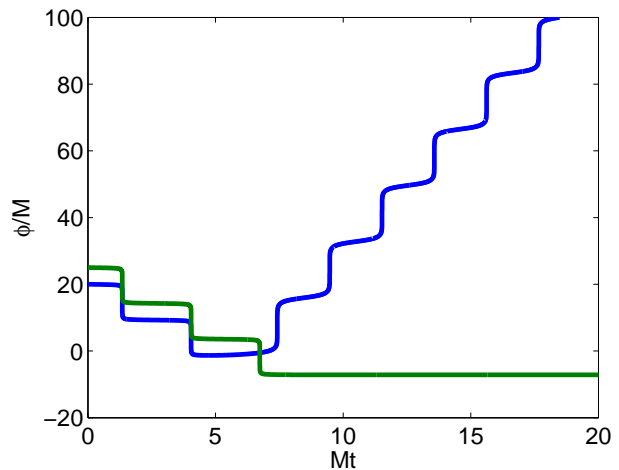}
\caption{\label{phi_pic}
Evolution of the field as a function of time with different initial conditions.  The evolutions correspond to the two cases depicted in Fig.~\ref{evolution}. During the turnover (classical) phase, the field remains practically constant since the kinetic energy is very small. In the bounce (quantum) phase the field rolls very rapidly. }
\end{figure}

A second important parameter determining the success of the exit is the ``phase'' of the universe when it encounters the steep potential incline, see Fig. \ref{evolution}. If it is in a phase of expansion, then $\phi$ slows down and the universe may not be able to make it to the top of the incline, and this is what happens in Fig. \ref{evolution}, the top panel. On the other hand, if it is in a contracting phase, then it is easier for $\phi$ to gain enough momentum to rise to the top of the hill. Fig.~\ref{phi_pic}, describes the evolution of the field during the cyclic and inflationary phases.

A third crucial parameter  is the hierarchy between the scale of the potential (which may be identified with the string or the inflationary scale) and the scale of the bounce (which we have assumed to be close to the Planck scale). As expected, the larger the hierarchy, the greater the probability of making the transition, because the scalar field gets that much more time to gain sufficient kinetic energy to cross the hill.

Finally, the steepness of the incline also plays an important role. If the rise is too shallow, the field may not be able to cross the whole way in one bounce, while if the rise is very sharp the force $V'(\phi)$ may overwhelm the ``anti-friction'' $3H\dot{\phi}$ in the contracting phase. Either case results in an unsuccessful exit, but the height of the vertical ``jumps'' in Fig.~ \ref{phi_pic}  increases with $\rho_\bullet$, and in fact for typical parameter values these jumps become much longer than the width of the slope, $\Lambda/\mu^3$. For instance, if one assumes both $\Lambda,\mu$ to be given by the string scale $M$ and $\rho^\frac{1}{4}_\bullet \sim M_{p} \sim 10^3 M$, numerical tests confirm that the phase transition takes place with significant probability, and is the typical outcome.

\section{Exploring the Landscape}
\label{landscape}

 We have seen how successfully the universe can overcome the potential barrier in going from the negative to the positive energy phase, but the potential in the negative region is not expected to have a monotonically inclining potential as $\phi\ra \infty$. Rather in a realistic setting one expects a varied landscape with local minima's and maxima's~\cite{Douglas:2006es}.

\begin{figure}
\includegraphics[width=8.75cm]{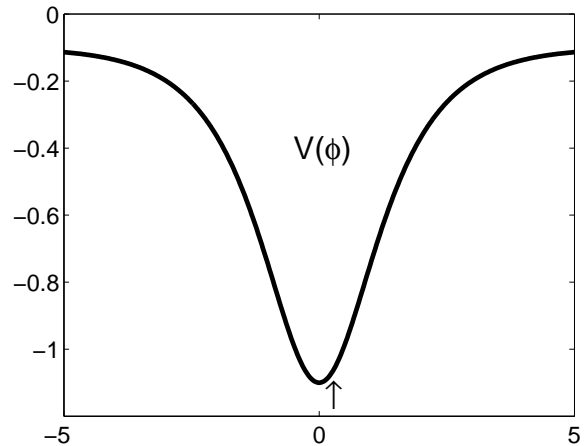}
\caption{\label{pot}
A plot of the potential $V(\phi)=-\Lambda(1+\sa\sec{\phi/M})$ when $\Lambda=0.1$ and $\sa=1$ in arbitrary units.  $\rho_\bullet=10^8$ in the plot.}
\end{figure}
\begin{figure}
\includegraphics[width=8.75cm]{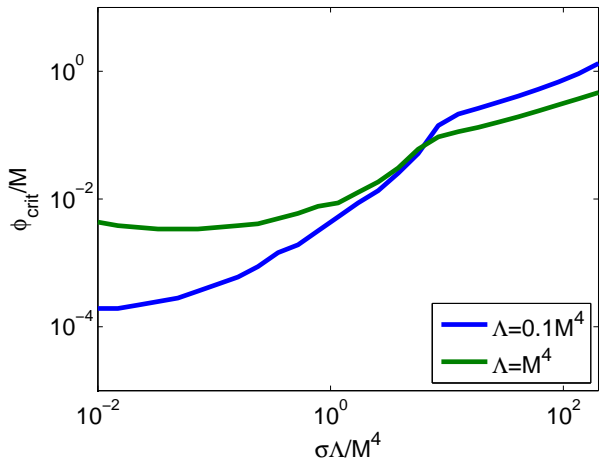}
\caption{\label{phi}
The field is released at rest. The critical value below which the field cannot climb up from the potential well is plotted as a function of the depth of the well.}
\end{figure}

One might wonder how efficient is the cyclic phase in glossing over these local minima's and continuing on in it's journey ``seeking'' the positive region?  We note that such a possibility was considered, though less quantitatively and in an anthropic context, in Ref. \cite{Piao:2004me}.
Here we explore this using the following potential, depicted in fig. \ref{pot}, as a prototype:
\be
V(\phi)=-\La\lp 1 + \sigma\mx{sech}(\phi/M)\rp.
\ee
As expected, the scalar field can easily gloss over such potential wells.  Now, we need not even consider any initial velocity, the field can gain enough kinetic energy as it approaches the minimum. In expanding cosmology the friction would of course hinder the field from climbing back the equal amount, but even if the turnaround occurs very close to the minimum,  the contraction can make it jump past the well. The definition of ``past the well'' is that $|\phi| > 2\sa^\frac{1}{3}$. We demonstrate this in Fig.~\ref{phi}, which shows the critical value of the field as a function of the potential depth $\sa$. If we release the field below this value, it doesn't climb over the well but can get stuck in the minimum. From these results it is clear that the field can naturally gloss over several minima in the negative potential region in it's quest to find a positive potential region, thereby making the exit scenario considerably more robust.

\section{Conclusion and discussion}\label{sec:conclusions}
{\bf Conclusions:}
To summarize, we have shown that a viable cosmology can be constructed from a landscape of $-$ve vacuum energy density. Such vacua are naturally expected in
string theory. This opens a new possibility that {\it our} patch of the universe could have begun with a negative cosmological constant. In particular, we focussed on a new class of inflationary models involving cyclic cosmology where entropy production in the cyclic phase make the cycles asymmetric and in fact the space-time starts resembling that of inflation over a longer time scale. Our results demonstrates that a graceful exit from the ``cyclic inflationary'' phase may indeed be possible lending viability to these new cosmological scenarios with very distinctive cosmic signatures~\cite{Biswas:2010si}.  It may also be possible to explain the smallness of the current positive cosmological constant in such scenarios~\cite{Prokopec:2006yh,Prokopec:2011ce}.

It is worth emphasizing though that the graceful exit mechanism discussed here can simply be a vehicle for the universe to jump from negative to positive energy phase, where traditional inflationary cosmology can take place.  As an added bonus the pre-inflationary cyclic phase makes the space-time ``past geodesically complete''~\cite{Borde,Linde}; in the infinite past the space undergoes quasi-periodic oscillations, the scale-factor never vanishes, and the particle trajectories can be extended to the infinite past.

\bibliography{lambdarefs}

\end{document}